\documentclass[superscriptaddress,showpacs,aps,twocolumn,prb,floatfix,longbibliography]{revtex4-2}
\usepackage{bm,amsmath,amssymb}
\usepackage{amssymb}
\usepackage{amsmath}
\usepackage{commath}
\usepackage{graphicx,bm}
\usepackage{verbatim}
\usepackage[latin1]{inputenc}
\usepackage{xcolor}
\usepackage{ulem}
\usepackage[colorinlistoftodos]{todonotes}

\presetkeys{todonotes}{inline}{}

\begin{document}

\title{Nonlocal thermoelectricity in quantum wires as a signature of Bogoliubov-Fermi points}

\author{Juan Herrera Mateos}
\affiliation{ECyT-ICIFI Universidad Nacional de San Mart\'{\i}n, Campus Miguelete, 25 de Mayo y Francia, 1650 Buenos Aires, Argentina}
\author{Leandro Tosi}
\affiliation{Grupo de Circuitos Cu\'anticos, Div. Dispositivos y Sensores, Centro At\'omico Bariloche (8400), San Carlos de Bariloche, Argentina}
\author{Alessandro Braggio}
\affiliation{NEST, Istituto Nanoscienze-CNR and Scuola Normale Superiore, I-56126 Pisa, Italy}
\author{Fabio Taddei}
\affiliation{NEST, Istituto Nanoscienze-CNR and Scuola Normale Superiore, I-56126 Pisa, Italy}
\author{Liliana Arrachea}
\affiliation{ECyT-ICIFI Universidad Nacional de San Mart\'{\i}n, Campus Miguelete, 25 de Mayo y Francia, 1650 Buenos Aires, Argentina}
\affiliation{Centro At\'omico Bariloche and Instituto de Nanociencia y Nanotecnolog\'{\i}a CONICET-CNEA (8400), San Carlos de Bariloche, Argentina}

\begin{abstract} 
 We study nonlocal thermoelectricity in a superconducting wire subject to spin-orbit coupling and a magnetic field with a relative orientation $\theta$ between them. We calculate the current flowing in a normal probe attached to the bulk of a superconducting wire, as a result of a temperature difference applied at the ends of the wire. 
 We focus on the linear response regime, corresponding to a small temperature bias.
 We find that the nonlocal thermoelectric response is strongly dependent on the angle $\theta$ and occurs in ranges which correspond to the emergence of
Bogoliubov Fermi points in the energy spectrum of the superconducting wire. 
\end{abstract}

%\pacs{ }
\maketitle

\section{\label{sec:intro}Introduction}
Superconducting quantum wires have garnered significant attention in various research fields, such as materials science, quantum physics, and condensed matter physics. 
 The appealing features of these systems rely on three crucial ingredients: resilient induced superconductivity,
%\cite{}, 
strong spin-orbit coupling (SOC)
%\cite{} 
and large gyromagnetic factor.
%\cite{}. 
%Besides the possible existence of Majorana zero modes, 
The goal of achieving a topological phase featuring Majorana zero modes \cite{wires1,wires2} was the driving force for numerous theoretical and experimental studies
into superconducting InAs wires \cite{mourik2012signatures,deng2016majorana,chen2017experimental,PhysRevLett.123.107703,
nichele2017scaling,vaitiekenas2020fullshell,alicea2012new,prada2020andreev,flensberg2021engineered}.
An equally fascinating phenomenon is the emergence of Bogoliubov-Fermi surfaces, whose signatures have been 
recently observed in InAs two-dimensional systems with an applied in-plane magnetic field proximitized by superconductors \cite{phan2022detecting}. The energetic stability and the topological properties of this peculiar phase have been the motivation of several theoretical studies  \cite{liu2003interior,wu2003superfluidity,forbes2005stability,Agterberg2017Mar,setty2020bogoliubov,Dutta2021Sep,Pal2024May,Ohashi2023Nov}.

In this paper, we show that the emergence of Bogoliubov Fermi points in a superconducting wire with SOC and magnetic field can lead to a strong nonlocal thermoelectric signature. 
These wires exhibit a topological phase across a range of chemical potentials ($\mu$), pairing amplitudes ($\Delta$) and Zeeman energies ($\Delta_B$)
subject to the condition that the  angle ($\theta$) between the directions of the SOC   and
the magnetic field satisfies $|\cos(\theta)|<\Delta/\Delta_B<1$ \cite{rex2014tilting,osca2014effects,klinovaja2015fermionic,aligia2020tomography,daroca2021phase}. 
Bogoliubov Fermi points emerge as the gap in the spectrum of the topological phase is partially closed by a twist beyond the critical angles defined by this condition.
This non-local thermoelectric response bears similarities to that 
recently proposed to take place in Josephson junctions of two-dimensional topological insulators
\cite{blasiPhysRevLett124,blasi2020nonlocal,blasi2021nonlocal}. 
In that case, such an intriguing effect is rooted in the helical nature of the Kramers pairs of edge states present in the system and
 it is induced by a Doppler shift generated by a magnetic flux threading the junction \cite{Tkachov2015Jul}. 
Instead,  in the case of the quantum wire, the pivotal role is played by the twist of the magnetic field giving rise to the Bogoliubov Fermi points.

\begin{figure}
	\centering
	\includegraphics[width=\columnwidth]{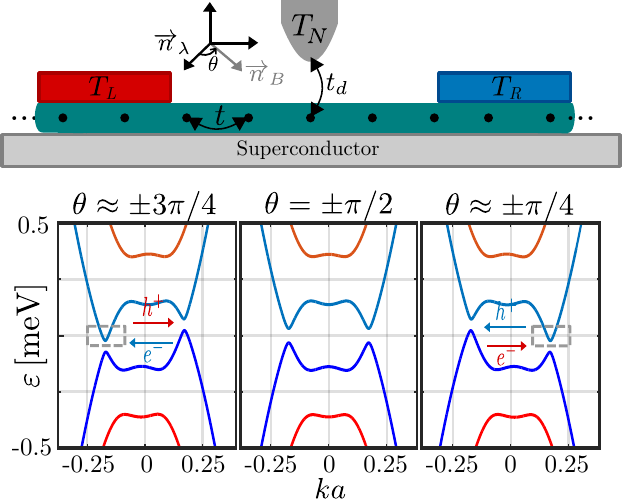}
	\caption{Top: Sketch of the setup.  Bottom: Spectrum of the Bogoliubov-de-Gennes Hamiltonian describing the wires with $t=5\,\rm{meV}$, $\Delta_B=0.5\,\rm{meV}$, $\lambda=0.25\,\rm{meV}$, $\Delta =0.2\,\rm{meV}$ and
 $\mu=-9.9\,\rm{meV}$,  for different values of the angle $\theta$ between the direction of the spin-orbit coupling (SOC) and the magnetic field. For $\theta=\pi/2$ the spectrum is fully gapped with two cones symmetrically aligned to $k=0$. For $0<\theta<\pi/2$ ($-\pi/2<\theta<0$) the cone with $k>0$  ($k<0$) crosses zero energy, defining 
 Bogoliubov Fermi points with right-moving electrons and left-moving holes (right-moving electrons and left-moving holes). These states account for the nonlocal thermoelectric response.}
	\label{fig:fig0}
\end{figure}

We consider the setup sketched in the top panel of Fig. \ref{fig:fig0}, where a quantum wire is proximitized with local s-wave superconductivity and has SOC and magnetic field acting in the directions $\vec{n}_\lambda$ and $\vec{n}_B$, respectively, with $\vec{n}_\lambda \cdot \vec{n}_B=\cos(\theta)$. A temperature difference ($T_L \neq T_R$) is imposed between the left ($L$) and the right ($R$) portions of the wire. A third terminal consisting of a normal-metal probe ($N$) is contacted at some point along the length of the wire with a 
tunnel-coupling $t_{\rm d}$. The nonlocal thermoelectric effect corresponds to an electrical current $J^{e}$ generated at the normal probe as a response to the transversal thermal bias.
We study this effect in the linear response regime, corresponding to a small temperature bias $T_L-T_R$.

The paper is organized as follows. The analysis of the spectral properties of the wire is presented in Sec. II. Sec. III is devoted to the evaluation of the current in the normal terminal and the definition of the local and nonlocal thermoelectric coefficients. Results are presented in the linear regime in Section IV. Sec. V is devoted to conclusions. Some technical details are explained in appendixes. 

\section{\label{sec:model}Spectral properties of the wire}
The wire is described by the Hamiltonian
$H_{\rm w}=(\frac{1}{2})\sum_k {\bf c}_k^{\dagger} {\cal H}_k,
{\bf c}_k$, which is expressed in the Nambu basis ${\bf c}_k=(c_{k \uparrow}, c_{k \downarrow},c^{\dagger}_{-k \downarrow},-c^{\dagger}_{-k \uparrow})^T$ and the Bogoliubov-de-Gennes (BdG) Hamiltonian matrix is given by \cite{wires1,wires2}
\begin{equation}\label{hbdg}
    {\cal H}_k= \tau^z \otimes \left[\xi_k \sigma^0 -  \lambda_k \vec{n}_{\lambda} \cdot \vec{\sigma} \right] -  \Delta_B \tau^0 \otimes \; \vec{n}_{\rm B} \cdot
    \vec{\sigma} + \Delta \tau^x \otimes \sigma^0.
\end{equation}
The Pauli matrices $\vec{\sigma}=\left( \sigma^{x}, \sigma^{y},\sigma^{z}\right) $ and $\vec{\tau}=\left( \tau^{x},\tau^{y},\tau^{z}\right) $ act in the spin and particle-hole degrees of freedom, respectively, while $\sigma^0, \; \tau^0$ are the identity
%2$\times$2 unitary 
 matrices. $\xi_k = - 2 t \cos (k a) - \mu$ is the kinetic dispersion relation relative to the chemical potential $\mu$, where $t$ is the nearest-neighbor hopping and $a$ is the lattice constant. The SOC is described by  $\lambda_k=  2 \lambda \sin (k a)$, while $\Delta_B=g \mu_B B$ is the Zeeman splitting due to the magnetic field $B$  and $\Delta$ is the local $s$-wave pairing potential.

In the results shown hereafter, we consider parameters of this Hamiltonian that are representative of reported experimental research in InAs wires \cite{deng2016majorana,chen2017experimental}. We assign $t=5\,\rm{meV}$, which fits the continuum model for the wires for
 $a=\sqrt{\hbar^2/2mt} \approx  15\,\rm{nm}$. We consider $\lambda=0.25\,\rm{meV}$ for the SOC, $\Delta=0.2\,\rm{meV}$ for the pairing potential
and a $g$-factor $g=18$, which corresponds to a Zeeman splitting energy $\Delta_B=0.5\,\rm{meV}$ for a magnetic field $B\approx 0.48\,{\rm T}$.

The eigenspectrum corresponding to the Hamiltonian of Eq.~(\ref{hbdg}) is shown in the bottom panel of Fig.~\ref{fig:fig0} for different values of $\theta$. We can identify two bands generated by the Zeeman splitting, which are doubled in the BdG representation. When the magnetic field is perpendicular to the SOC ($\theta=\pm\pi/2$) the spectrum is fully gaped for all values of $k$. Due to the combination of the SOC and $B$, the effective pairing has s-wave as well as p-wave components \cite{wires1,wires2,tomoPhysRevLett.125.256801,grun}. The latter is the dominant one when the system is in the topological phase for $ 0 \leq \mu +2t \leq \sqrt{\Delta_B^2-\Delta^2}$. 
%This happens when the Fermi energy of the wire without superconductivity is within the gap generated by the magnetic field. 
This is precisely the situation illustrated in the figure. 
%It is important to notice that the effective gap is significantly smaller than the amplitude of the pairing potential $\Delta$. 
When the orientation of the magnetic field is twisted, such that $\theta$ overcomes the critical values defined by the condition $|\cos(\theta)|<\Delta/\Delta_B<1$,
 the superconducting gap is partially closed. In fact, the ``cones'' of the spectrum cross zero energy from positive (negative) energies defining Bogoliubov-Fermi points for $k>0$ ($k<0$). 
 %These are the 1D counterpart of the Bogoliubov-Fermi surfaces emerging in 2D superconductors with SOC when an in-plane magnetic field is applied (see Ref. \cite{phan2022detecting}). 
 The right (left) bottom panels of Fig. \ref{fig:fig0} correspond to $\theta=\pm \pi/4$. 
 
The aim of this paper is  to show that the scenario of Bogoliubov-Fermi points illustrated in Fig. \ref{fig:fig0} hosts the fundamental ingredients to have a nonlocal thermoelectric response. It is well known that a necessary condition for the phenomenon of thermoelectricity to take place for the transmission probabilities not to be even in energy
\cite{benenti2017fundamental}. This condition usually relies on the implementation of energy filters in two-terminal configurations and is difficult to realize in superconductors since these systems are intrinsically particle-hole symmetric \cite{ozaeta2014predicted,kolenda2017thermoelectric,shapiro2017thermoelectric,keidel2020demand,mukhopadhyay2022thermal,machon2013nonlocal,mazza2015separation,heidrich2019nonlocal,PhysRevResearch.6.L012049,PhysRevB.106.115419}.
In fact, we see that the three spectra shown in Fig. \ref{fig:fig0} have this symmetry. The key ingredient for nonlocal thermoelectricity in the setup we are studying is to generate an imbalance between left-moving electrons (thermalized with the right reservoir) and right-moving holes (thermalized with the left reservoir). Hence, as a consequence of an applied temperature difference at the superconducting reservoirs, the fluxes associated with the two types of quasiparticles into the normal probe are not compensated and a net current is generated. In the spectrum of Fig. \ref{fig:fig0} with  $\theta=\pm \pi/2$ the low-energy cone with $k>0$ ($k<0$) corresponds to a right-moving electron (hole) and a left-moving hole (electron). Importantly, the spectrum is symmetrical to $k=0$, implying identical velocities and densities of states of the left and right movers. In the twisted case, we can identify a low-energy branch of bgoliubons forming Fermi points with electrons moving to the right and holes moving to the left (see plots with  $\theta=\pm \pi/4$). The opposite situation takes place for $\theta=\pm 3 \pi/4$. This mechanism may display a thermoelectric response since it produces the necessary particle-hole imbalance. 

In Sec. IV, we show explicit calculations of the thermoelectric current that confirm this picture. It is interesting to compare with the situation discussed in Ref. \cite{blasiPhysRevLett124} for a device where the Kramers pair of helical edge states of a topological insulator in a Josephson junction.
%between two superconductors. 
In that case, the imbalance between electrons and holes was induced by a Doppler shift generated by the magnetic flux threading the junction. Although different, both systems share common features. In fact, in both cases, the low-energy spectrum hosts a pair of
left-right movers with different spin orientations in contact with a $s$-wave superconductor. Because of the broken
SU(2) symmetry, in both systems superconducting pairing is induced in both $s$-wave and $p$-wave channels. The effect of the twisted magnetic field in our case and the Doppler shift in the case of Ref. \cite{blasiPhysRevLett124} is to introduce asymmetry in the spectrum so that a single pair of particle-hole quasiparticles moving into opposite directions dominate the quantum energy transport.

\section{\label{sec:current}Thermoelectric transport}

 We now present the theoretical approach to calculate the expression for the current in terms of  non-equilibrium Green's function formalism
 (see Refs.\cite{Cuevas1996Sep,Alvarado2020Mar,grun}). 

\subsection{Model of the device}

The full Hamiltonian reads
\begin{equation}\label{hsn}
    H=\frac{1}{2}\left[H_{\rm w}+ H_{\rm d} + H_{\rm N} + H_{\rm cont}\right],
\end{equation}
where the Hamiltonian for the superconducting wire $H_{\rm w}$ is the same defined in Eq. (\ref{hbdg}). It is convenient here to express it in real space as follows
\begin{widetext}
%\begin{eqnarray}\label{hsreal}
%& & H_{\rm w}  =  - \sum_{j=-\infty}^{\infty} %\left[   {\bf c}^{\dagger}_{j} \tau^z \otimes %\left( t_j \sigma^0+ i \lambda_j \; %\vec{n}_{\lambda} \cdot \vec{\sigma}  \right) {\bf %c}_{j+1} + \text{H.c.} \right] + \nonumber \\
%&  &\sum_{j=1}^{\infty} {\bf c}^{\dagger}_{j} %\left[\Delta \tau^x \otimes \sigma^0 - B_j \tau^0 %\otimes \vec{n}_{B} \cdot \vec{\sigma}  -\mu_j %\tau^z \otimes \sigma^0  \right]{\bf c}_{j},
%\end{eqnarray}
\begin{equation}\label{hsreal}
 H_{\rm w}  =  - \sum_{j=-\infty}^{\infty} \left[   {\bf c}^{\dagger}_{j} \tau^z \otimes \left( t \sigma^0+ i \lambda \; \vec{n}_{\lambda} \cdot \vec{\sigma}  \right) {\bf c}_{j+1} + \text{H.c.} \right] +
\sum_{j=-\infty}^{\infty} {\bf c}^{\dagger}_{j} \left[\Delta \tau^x \otimes \sigma^0 - \Delta_B \tau^0 \otimes \vec{n}_{B} \cdot \vec{\sigma}  -\mu \tau^z \otimes \sigma^0  \right]{\bf c}_{j},
\end{equation}
\end{widetext}
with ${\bf c}_{j} =\left( c_{j,\uparrow}, c_{j,\downarrow},c^{\dagger}_{j,\downarrow}, -c^{\dagger}_{j,\uparrow} \right)$. 

The Hamiltonian for the normal probe is a one-dimensional tight binding Hamiltonian with hopping $t_{\rm N}$,
\begin{eqnarray}\label{hn}
    H_{\rm N} &=& -t_{\rm N}
\sum_{j=1}^{\infty} 
%\left[ 
\left({\bf b}_{j}^{\dagger } \tau^z \otimes \sigma^0
{\bf b}_{j+1}+\text{H.c.}\right) 
%\right. \nonumber \\& &  \left. -  \; {\bf b}_{j}^{\dagger } \left(\mu_{\rm N} \tau^z \otimes \sigma^0+ B_{\rm N} \tau^0 \otimes \vec{n}_{B} \cdot \vec{\sigma} \right){\bf b}_{j} \right],
\end{eqnarray}
where we are using the notation ${\bf b}^{\dagger}_j=\left(b^{\dagger}_{j,\uparrow},
b^{\dagger}_{j,\downarrow},b_{j,\downarrow},- b_{j,\uparrow} \right)$ for the Nambu spinor within the normal lead. This system is assumed to
be at temperature $T_N$ and voltage $V$. 
 The interface is modeled by an intermediate site d, which plays the role of a quantum dot. 
 %and Anderson impurity model
\begin{eqnarray}\label{hd}
%    H_{\rm d}&=&H_{\rm d,0}+ U n_{d\uparrow} n_{d\downarrow}, \nonumber \\
   H_{\rm d} &=& - {\bf d}^{\dagger} \varepsilon_{\rm d} \tau^z \otimes \sigma^0 
   {\bf d},
\end{eqnarray}
where ${\bf d}=\left(d_{\uparrow}, d_{\downarrow},
d^{\dagger}_{\downarrow}, -d^{\dagger}_{\uparrow}\right)^T$ is the Nambu spinor that describes the degrees of freedom in the quantum dot and
%$U$ the Coulomb repulsion, 
$\varepsilon_{d}$ is a local energy representing a barrier.

The last term of Eq. (\ref{hsn}) is the tunneling-contact between the quantum dot and the wire and the normal probe. It reads
\begin{equation}\label{hcont}
    H_{\rm cont} = -
%\sum_{s=\uparrow, \downarrow} \left[\left( t_{\rm cS} c^{\dagger}_{1,s} \; + \;  t_{\rm cN} b^{\dagger}_{1,s} \right) d_s + \text{H. c.} \right],
\left[\left( t_{\rm d} {\bf c}^{\dagger}_{0} \; + \;  t_{\rm N} {\bf b}^{\dagger}_{1} \right) \tau^z \otimes \sigma^0 {\bf d} + \text{H. c.} \right],
\end{equation}
where the label $\ell=0,1$ denotes the sites of the wire and normal chains that are tunnel-coupled to the interface, respectively. Notice
that for $\varepsilon_{\rm d}=0$ the quantum dot is assimilated to the normal lead. 

 In the  calculation, we split the  wire into a central segment 
containing $N_{\rm w}$ lattice sites, which is contacted to left ($L$) and right ($R$) to semi-infinite wires described by the same lattice Hamiltonian. These play the
role of reservoirs with 
temperatures 
$T_L$ and $T_R$, respectively.

\subsection{Current in the normal lead}
The current flowing between the connecting site $d$ and the normal lead reads 
\begin{equation}  \label{current}
    J^e= \frac{e }{h}\mbox{Re} \{ \int  d\varepsilon \mbox{Tr} \left[\tau^z \otimes \sigma^0 \; {\bf t}_{\rm N}{\bf G}^<_{Nd}(\varepsilon) \right]\},
\end{equation}
where ${\bf t}_{\rm N}= t_{\rm N} \tau^z \otimes \sigma^0$ and we have introduced the 
lesser Green's function
\begin{equation}
 {\bf G}^<_{Nd}(t,t^{\prime}) =-i\langle {\bf d}^{\dagger}(t^{\prime}) {\bf b}_1(t) \rangle,
\end{equation}
as well as the  Fourier transform $t-t^{\prime} \rightarrow \varepsilon$. 

Using properties of the Green's functions presented in Appendix \ref{sec:GreenCalculations} and following the details developed in Appendix \ref{sec:DetailsOfJ}, the current can be expressed as follows,
%\begin{equation}\label{curf}
%    J=\frac{e }{2h}\int  d\varepsilon %\left[\sum_{j=L,R} \left( f_j -f^+_N\right)%{\cal T}_j(\varepsilon) + 
%    \left[f^-_N  -f^+_N  \right] {\cal %R}_A(\varepsilon)\right] \\ 
%\end{equation}
\begin{eqnarray}
%\begin{split}
J^e&= & \frac{e }{2h}\int  d\varepsilon \left\{\sum_{j=L,R} \left [\left( f_j -f^+_N\right){\cal T}_j^{(p)}(\varepsilon) \right.\right.\\
& & \left. \left. - \left( f_j -f^-_N \right){\cal T}_{j}^{(h)}(\varepsilon)  \right ]+ 2\left[f^-_N  -f^+_N  \right] {\cal R}_A(\varepsilon)\right\}.\nonumber 
   \label{corriente1}
%\end{split}
\end{eqnarray}
%\begin{equation}
%\begin{split}
%J^e= & \frac{e }{2h}\int  d\varepsilon \left\%{\sum_{j=L,R} \left [\left( f_j -f^+_N\right){\cal T}_j^{(p)}(\varepsilon) \right.\right.\nonumber \\
%&  \left. \left. \;\;\;\;\; \;\;\;\;\; - \left( f_j -f^-_N \right){\cal T}_{j}^{(h)}(\varepsilon)  \right ]\right\} \\ & +\frac{e }{h}\int  d\varepsilon
%   \left[f^-_N  -f^+_N  \right] {\cal %R}_A(\varepsilon),
%   \label{corriente1}
%\end{split}
%\end{equation}
The first terms describe the normal transmission for the particle ($p$) and holes ($h$),
and read
%\begin{widetext}
\begin{equation}
\label{tpj}
{\cal T}^{(p)}_j(\varepsilon) = \sum_{\ell=1,2} \left\{\boldsymbol{\Gamma}_{N}(\varepsilon)
%\left[ \boldsymbol{\Sigma}^r_N - \boldsymbol{\Sigma}^a_N\right]
{\bf \cal G}^r_{dj}(\varepsilon) \boldsymbol{\Gamma}_{j}(\varepsilon)
%\left[\boldsymbol{\Sigma}^a_{j}(\varepsilon)-\boldsymbol{\Sigma}^r_{j}(\varepsilon) \right]
{\bf \cal G}^a_{jd}(\varepsilon)\right\}_{\ell,\ell},\nonumber
\end{equation}
\begin{equation}\label{thj}
{\cal T}^{(h)}_j(\varepsilon) = \sum_{\ell=3,4} \left\{\boldsymbol{\Gamma}_{N}(\varepsilon)
%\left[ \boldsymbol{\Sigma}^r_N - \boldsymbol{\Sigma}^a_N\right]
{\bf \cal G}^r_{dj}(\varepsilon) 
\boldsymbol{\Gamma}_{j}(\varepsilon)
%\left[\boldsymbol{\Sigma}^a_{j}(\varepsilon)-\boldsymbol{\Sigma}^r_{j}(\varepsilon) \right]
            {\bf \cal G}^a_{jd}(\varepsilon)\right\}_{\ell,\ell},
\end{equation}
%where $\boldsymbol{\Gamma}_{j}(\varepsilon)=\boldsymbol{\Sigma}^a_{j}(\varepsilon)-\boldsymbol{\Sigma}^r_{j}(\varepsilon)$, with $j=L,R,N$, are the hybridization matrices defined in Appendix \ref{sec:GreenCalculations}.
%\end{widetext}
%\begin{eqnarray}\label{tpj}
%   & &  {\cal T}^{(p)}_j(\varepsilon) = \\
%   & & \sum_{\ell=1,2} \left\{\left[ %\boldsymbol{\Sigma}^r_N - \boldsymbol{\Sigma}^a_N\right]{\bf \cal G}^r_{dj}(\varepsilon) \left[\boldsymbol{\Sigma}^a_{j}(\varepsilon)-\boldsymbol{\Sigma}^r_{j}(\varepsilon)
%            \right]{\bf \cal G}^a_{jd}%(\varepsilon)\right\}_{\ell,\ell},\nonumber
%\end{eqnarray}
%\begin{eqnarray}\label{thj}
%   & &  {\cal T}^{(h)}_j(\varepsilon) = \\
%   & & \sum_{\ell=3,4} \left\{\left[ \boldsymbol{\Sigma}^r_N - \boldsymbol{\Sigma}^a_N\right]{\bf \cal G}^r_{dj}(\varepsilon) \left[\boldsymbol{\Sigma}^a_{j}(\varepsilon)-\boldsymbol{\Sigma}^r_{j}(\varepsilon)
%            \right]{\bf \cal G}^a_{jd}(\varepsilon)\right\}_{\ell,\ell},\nonumber
%\end{eqnarray}
while the last term of Eq. (\ref{corriente1}) describes the Andreev reflection and reads,
\begin{equation}
{\cal R}_A(\varepsilon) =\sum_{\ell=1,2,\overline{\ell}=3,4} \boldsymbol{\Gamma}_{N}(\varepsilon)
%    \left[ \boldsymbol{\Sigma}^r_N - \boldsymbol{\Sigma}^a_N\right]_{\ell,\ell}
    {\bf G}^r_{dd}(\varepsilon)_{\ell,\overline{\ell}} \boldsymbol{\Gamma}_{j}(\varepsilon)
%    \left[ \boldsymbol{\Sigma}^a_N -\boldsymbol{\Sigma}^r_N \right]_{\overline{\ell},\overline{\ell}} 
 {\bf G}^a_{dd}(\varepsilon)_{\overline{\ell},\ell}. 
\end{equation}
The functions ${\cal T}^{\rm (p)}_j(\varepsilon)$ and ${\cal T}_j^{\rm (h)} (\varepsilon)$ are, respectively, the transmission probabilities for an electron-like and hole-like quasiparticle starting from the superconducting lead $j=L,R$ to go in lead $N$, while ${\cal R}_A(\varepsilon)$ is the Andreev reflection probability for an electron starting from lead $N$ to be reflected as a hole.

%\begin{eqnarray}
%   & &  {\cal R}_A(\varepsilon) = \\
%  & & \sum_{\ell=1,2,\overline{\ell}=3,4} 
%    \left[ \boldsymbol{\Sigma}^r_N - \boldsymbol{\Sigma}^a_N\right]_{\ell,\ell}
%    {\bf G}^r_{dd}(\varepsilon)_{\ell,\overline{\ell}}\left[ \boldsymbol{\Sigma}^a_N -\boldsymbol{\Sigma}^r_N \right]_{\overline{\ell},\overline{\ell}} 
 %{\bf G}^a_{dd}(\varepsilon)_{\overline{\ell},
 %\ell},\nonumber
%\end{eqnarray}
%where the contributions of elements 1 and 2, and 3 and 4 are equal.
We have introduced the hybridization matrices $\boldsymbol{\Gamma}_{j}(\varepsilon)$, the non-local retarded/advanced Green's functions
${\bf \cal G}^{r/a}_{jd}(\varepsilon)$, with $j=L,R,N$, as well as the local ones ${\bf G}^a_{dd}(\varepsilon)$. The calculation of all these quantities is explained
in Appendix \ref{sec:GreenCalculations}.
We have also introduced the Fermi functions $f_j(\varepsilon)=1/(e^{\beta_j \varepsilon}+1)$
and $f_N^{\pm}(\varepsilon)=f_N(\varepsilon \mp eV)$,
 where $\beta_j$ is the inverse temperature of the reservoir $j$.
Notice that only the normal lead is biased with a voltage $V$.

Some properties of these transmission functions are discussed in Appendix \ref{sec:PropertiesTransmissionFunctions}.

\subsection{\label{sec:LinearResponseMAIN} Linear response}
%\subsection{Onsager coefficients}
We particularly focus on the linear-response regime and on a range of temperatures below the critical temperature of the superconductor. We consider the general case where the temperatures for a normal probe, and the left and right terminals of the wire are, respectively,
\begin{eqnarray} \label{temperatures}
T_N &=&T, \\
T_L&=&T+\frac{\Delta T}{2}, \;\;\;\;\;  T_R=T+r\,\frac{\Delta T}{2}, \;\;\;-1\leq r\leq 1\nonumber.
\end{eqnarray}
with $\Delta T\ll T$ being infinitesimal.
Notice that when $r=-1$ the temperature bias at the superconducting wire is perfectly symmetric with respect to the reference temperature of the normal probe. In the opposite case, where  $r=1$, only one of the terminals of the wire is thermally biased with respect to the normal probe.

We consider that  the ends of the wire are grounded ($\mu_L=\mu_R=0$), while the normal terminal may have an electrical bias $\mu_N=eV$. We define the affinities $X_V=V/T$ and $X_T= \Delta T/(2T^2)$ and assume they are small enough to justify treating them in the linear response.
Expanding the differences of Fermi functions entering Eq.~(\ref{corriente1}) up to first order in these affinities, we get
%, we have
%: $f_j-f_N^{\pm} \approx \pm df/(d\varepsilon)\,eV$ and $f_N^{-}-f_N^{+} \approx df/(d\varepsilon)\,2eV$. Similarly, expanding the Fermi functions up to the first order 
%$f_j-f_N^{\pm} \approx \pm df /(dT)\,\Delta T_j= -(\varepsilon/T)df/(d\varepsilon) \Delta T_j$, where $\Delta T_L=\Delta T/2$, while $\Delta T_R=r \Delta T/2$
%and $f_N^{-}-f_N^{+}=0$. 
%The induced charge current at the normal lead results in 
%\begin{equation}
%$J^e= {\cal L}_{ee} X_V+  {\cal L}^{\rm nl}_{eq} X_T$.
%\end{equation}
\begin{equation}
    J^e= {\cal L}_{ee} X_V + \left[ \frac{1+r}{2}{\cal L}_{eq}^{\rm loc} + \frac{1-r}{2}{\cal L}_{eq}^{\rm nl} \right]X_T.
\end{equation}

The Onsager coefficients read 
\begin{eqnarray} \label{onsa}
    {\cal L}_{ee} &=& -\frac{e^2\,T}{2h} \int d\varepsilon \left[{\cal{T}}_L^+(\varepsilon)+{\cal{T}}_R^{(+)}(\varepsilon)  + 4 {\cal R}_A(\varepsilon) \right]
   \frac{d f(\varepsilon)}{d\varepsilon},\,\nonumber \\ 
   {\cal L}_{eq}^{\rm loc} &=&-\frac{e T}{2h} \int d\varepsilon \left[{\cal T}^{-}_L(\varepsilon) +{\cal T}^{-}_R(\varepsilon) \right]
   \varepsilon \frac{d f(\varepsilon)}{d\varepsilon},\, \nonumber \\ 
   {\cal L}_{eq}^{\rm nl} &=&-\frac{e T}{2h}  \int d\varepsilon \left[{\cal T}^{-}_L(\varepsilon) -{\cal T}^{-}_R(\varepsilon) \right]
   \varepsilon \frac{d f(\varepsilon)}{d\varepsilon},\
\end{eqnarray}
where ${\cal T}^{\pm}_j(\varepsilon)={\cal T}_j^{\rm (p)} (\varepsilon)\pm {\cal T}_j^{\rm (h)} (\varepsilon)$. 
The Fermi function 
%$f(\varepsilon)=1/(e^{\varepsilon/k_B T}+1)$ 
is evaluated at the base temperature $T$. The derivative of this function entering the coefficients of Eq. (\ref{onsa}) defines the relevant transport window $|\varepsilon| \simeq k_B T$. %The $\pm$ signs entering the transmission functions of ${\cal L}_{ee}$ and ${\cal L}^{\rm nl}_{eq}$ reflect the fact that they describe transport generated by a voltage and temperature bias, respectively. 

Notice that ${\cal L}_{ee}$ and ${\cal L}_{eq}^{\rm loc}$ are local quantities, which corresponds to the convergence of the transport channels of the two superconducting terminals into the $N$ one. Instead ${\cal L}^{\rm nl}_{eq}$ is a nonlocal quantity, corresponding to the difference in the thermoelectrical transport between the $L$ and $R$ terminals and the $N$ one and we have stressed this property with the label ``nl''. 
In Appendix C we show that these functions satisfy 
${\cal T}_L^{\rm (p/h)}(\varepsilon,\theta)={\cal T}_R^{\rm (p/h)}(\varepsilon,\theta+\pi)$ which implies a change of sign of ${\cal L}^{\rm nl}_{eq}(\theta)$ at $\theta =\pm \pi/2$.

The relevant transport coefficients we discuss next are the conductance $G$ and the {\it nonlocal} ({\it local}) Seebeck coefficient $S^{\rm nl}$($S^{\rm loc}$). These are defined from the Onsager parameters as
\begin{equation}
G= \frac{{\cal L}_{ee}}{T}, \;\;\;\;\; S^{\rm nl/loc}= \frac{{\cal L}^{\rm nl/loc}_{eq}}{T {\cal L}_{ee}}.  
\end{equation}
 %The conductance is bounded by $4G_0$ and $2G_0$ {\color{red} (FT: true only for E mucu bigger tha Delta)} within the superconducting gap and above the superconducting gap, respectively, being $G_0=e^2/h$. 

\section{\label{sec:intro}Numerical Results}
In the calculations %\st{we consider the wire modeled by the Hamiltonian of Eq. (\ref{hbdg})} 
we evaluate the Green's functions of the semi-infinite wires representing 
the reservoirs at the temperatures $T_L$ and $T_R$  with a recursive method \cite{lopezsancho}.
In linear response, the results do not depend on either the length of the central wire or on the position of the contact with the normal probe, but depend on the tunneling coupling $t_{\rm d}$ between the normal probe and the wire. 
For simplicity, we consider  $t_{\rm d}= t_{\rm N}$ and $\varepsilon_{\rm d}=0$.  

\subsection{Non-local thermoelectric response}
\begin{figure}
	\centering
	\includegraphics[width=\columnwidth]{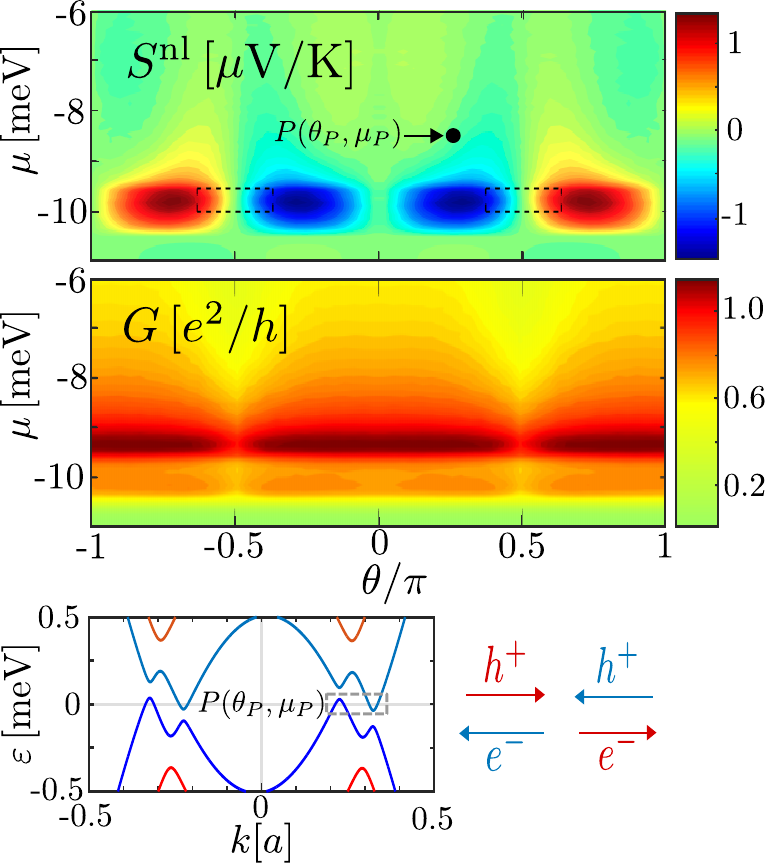}
	\caption{$S^{\rm nl}$ and $G$ as functions of $\theta$ and $\mu$ for the same parameters of the wire as in Fig. \ref{fig:fig0}. Other parameters are
 $t_d=2.5\,\rm{meV}$ and $k_BT/\Delta=0.25$. The topological phase is within the rectangles indicated in dashed lines. Bottom panel: spectrum corresponding to the point $(\theta_{P},\mu_{P})$ for which the nonlocal thermoelectric response is weak. } 
 \label{fig:fig2}
\end{figure}

In Fig. \ref{fig:fig2} we present the resulting conductance and Seebeck coefficient as functions of the chemical potential $\mu$ and relative SOC-magnetic field angle $\theta$. We can identify in the top panel of the figure, regions where the Seebeck coefficient takes large positive and negative values.
These are the nonlocal thermoelectric features anticipated from the analysis of the spectrum with Bologiubov Fermi points. In fact, notice that the value of $\mu$ corresponding to the spectrum of Fig. \ref{fig:fig0} is precisely the one for which the largest values of $S^{\rm nl}$ are achieved. We can verify the vanishing nonlocal thermoelectric response for $\theta=0, \pm \pi/2, \pm \pi$ for which the spectrum is symmetric with respect to $k=0$. In addition, the opposite signs of $S^{\rm nl}$ for given angles $\theta$ and $\theta+\pi$ are consistent with the interchange of right-moving particle-like and left-moving hole-like quasiparticles observed in the spectra of Fig. \ref{fig:fig0} and also with the symmetry properties of the functions ${\cal T}_L^{\rm (p/h)}(\varepsilon,\theta)$. In the bottom panel of Fig. \ref{fig:fig2} we show, for comparison the spectrum for the parameters $(\theta_{P},\mu_{P})$ indicated in the top panel where the thermoelectric response is weaker, albeit non-vanishing. 
For $\mu+2t \simeq   \sqrt{\Delta_B^2-\Delta^2}$  the dominant superconducting pairing is an s-wave type and four quasiparticle cones emerge at low energy in the spectrum (two for $k>0$ and two for $k<0$). For $\theta \neq \pi/2$, four Bogoliubov-Fermi points emerge at each side of $k=0$. Focusing at $k>0$, we can identify an electron-hole cone crossing zero energy from above along with a hole-electron cone crossing zero energy from below. The nature of these low-energy quasiparticles is consistent with a pair of left-moving electrons and right-moving holes partially compensated by a pair of right-moving electrons and left-moving holes. Consequently, there is a partial cancellation of the nonlocal thermoelectric transport. In conclusion, the nonlocal thermoelectrical effect is much stronger when a single  Bogoliubov-Fermi cone is present.
This is precisely the case for emerging Fermi points within the topological phase.

%The behavior of the conductance depends strongly on the density of states of the wire and on the coupling $t_{\rm d}$. 
The behavior of the conductance is affected by the density of states of the wire and the coupling $t_{\rm d}$ between wire and the normal probe.
It achieves a maximum close to $\mu+2t \sim   \sqrt{\Delta_B^2-\Delta^2}$, just above  the boundary for the topological phase. This is because the density of states of the wire is large for this value of $\mu$ and the two spin channels contribute to the transport.

\begin{figure}
	\centering
	\includegraphics[width=\columnwidth]{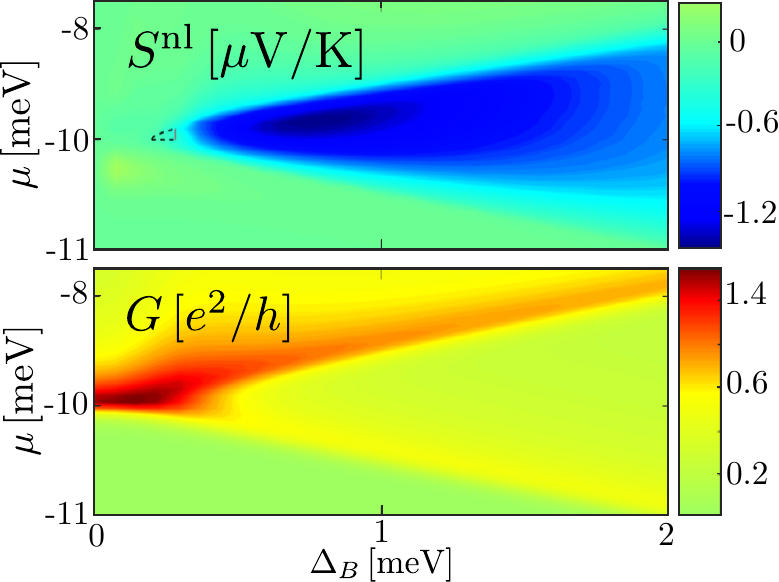}
	\caption{$S^{\rm nl}$ and $G$ as functions of $\Delta_B$ and $\mu$ for $\theta=\pi/4$. Other parameters are the same as the previous figures.
 The topological phase is inside the small triangle highlighted in a dashed line in the upper panel.
 It is defined by $0\leq \mu+2t\leq \sqrt{\Delta_B^2-\Delta^2}$ and $\abs{\cos(\theta)}\leq\Delta/\Delta_B$ (vertical gray line).
	\label{fig:ContornoByMu}
 }
 %\todo{
 %\textbf{AB:}Is it possible to get a dashed line for the topological phase condition in the bottom panel of Fig 3 to better identify the topological phase condition $0 \leq \mu +2t \leq \sqrt{\Delta_B^2-\Delta^2}$? I suspect that the lines which defines the maximal value of the nonlocal thermolectricity area are, in some measure connected to the topological phase. }
\end{figure}

A complementary and helpful perspective can be obtained by fixing the angle and changing the magnetic field. In Fig. \ref{fig:ContornoByMu} we focus on $\theta=\pi/4 $ and show again $S^{\rm nl}$ and $G$ as functions of $\Delta_B$ and $\mu$. In the behavior of these quantities we can identify the gap $\sim \Delta_B$ (see the blue region in the upper panel), within which $G$ is minimal while the nonlocal thermoelectric response is strongest.
The small triangle with the dashed line in the upper panel of this figure defines the boundary for the topological phase. As in the previous figure we see that the
maximal response in $S^{\rm nl}$ is associated with the emergence of the Bogoliubov-Fermi points. Such an effect occurs as the magnetic field is twisted beyond the critical value defined by $\abs{\cos(\theta)} < \Delta/\Delta_B$. From the experimental point of view, it is important to notice
that $S^{\rm nl}$ remains close to the maximal values across a wide range of $\mu$ and $\Delta_B$, which facilitates the exploration of this effect by varying the magnetic field. 
%there is a  where $S^{\rm nl}$ remains close to these maximal values which is relevant for experiments where the chemical potential is hard to tune (depends on sample growth method) but the in-plane magnetic field can be varied in a reasonable range.  }

\begin{figure}
	\centering
	\includegraphics[width=0.8\columnwidth]{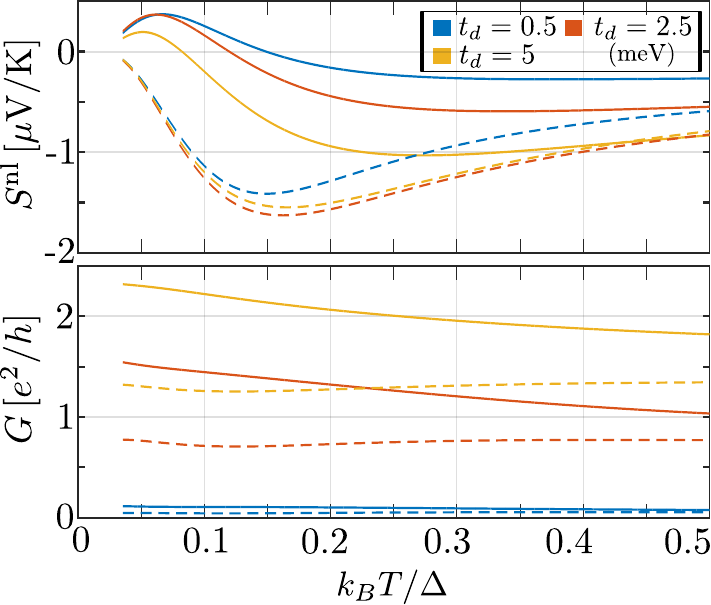}
	\caption{$S^{\rm nl}$ and $G$ as functions of temperature for the same parameters as the previous figures. 
 %$t=5\,\rm{meV}$, $\Delta_B=0.5\,\rm{meV}$, $\lambda=0.25\,\rm{meV}$, $\theta=\pi/4$, $\Delta=0.2\,\rm{meV}$ 
 and different values of coupling with the normal reservoir, $t_d$. 
 Solid (dashed) lines correspond to $\mu=-9.38\,m\rm eV$ ($\mu=-9.9\,m\rm eV$).
% The chemical potential was chosen from the maximum values of each coefficient as found in Fig. \ref{fig:fig2}; namely   (solid lines) for $S^{\rm nl}$ and  (dashed lines) for $G$.
}
	\label{fig:fig3}
\end{figure}

Finally, in Fig. \ref{fig:fig3} we show $S^{\rm nl}$ and $G$ obtained for different couplings $t_{\rm d}$ between the normal probe and the wire, as a function of the temperature $T$.  
We focus on low enough $T$, so that we can neglect dependence of   $\Delta$ on $T$. It is however easy to numerical introduce the self-consistent temperature correction of the gap in the case if needed.
The amplitude of the non-local Seebeck coefficient $S^{\rm nl}$ 
 decreases as a function of the temperature. This is consistent with the fact that, for increasing temperature, high-energy regions of the spectrum play a role. Such excitations 
 contain electrons and holes traveling in opposite directions, with the concomitant suppression of the non-local thermoelectric response. This thermoelectric behavior is anomalous. It strongly differs from the standard behavior of the Seebeck coefficient which typically scales with the temperature. The effect of temperature is clearly weaker in the conductance. In addition, the thermoelectric response is not strongly affected by the coupling $t_{\rm d}$, while the opposite is true regarding the conductance.

\subsection{ Non-local vs local thermoelectric response.}
Depending on how symmetrical the temperature difference between the superconducting reservoirs is, we have a non-local, local thermoelectric response or a combination of both. 

Given the temperatures as defined in Eq. (\ref{temperatures}), we see that the thermoelectric response is {\em purely nonlocal} when the temperature bias at the superconducting wire is perfectly symmetric with respect to the reference temperature of the normal probe, which corresponds to $r=-1$. In the opposite limit, where  $r=1$, the temperature bias is completely asymmetric, since only one of the terminals of the wire is thermally bias with respect to the normal probe, and in this limit, only the local component contributes. Intermediate 
situations correspond to $-1<r<1$ and the two components contribute.

\begin{figure}
	\centering
	\includegraphics[width=\columnwidth]{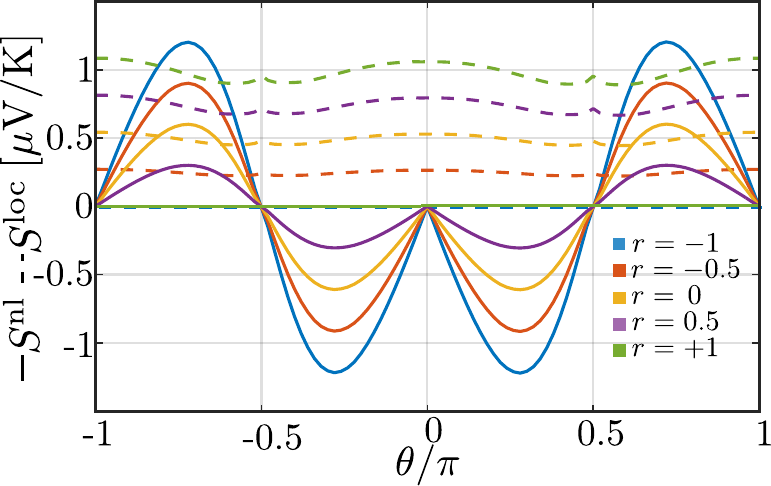}
	\caption{Non-local and local Seebeck coefficient as function of $\theta$, for a system with $t=5\,\rm{meV}$, $\Delta_B=0.5\,\rm{meV}$, $\lambda=0.25\,\rm{meV}$, $\Delta=0.2\,\rm{meV}$, $t_d=2.5\,\rm{meV}$, $\mu=-9.9\,{\rm{meV}}$, for different values of $r$, the factor that quantifies the asymmetry of temperatures between the superconducting reservoirs L and R. $r=-1$ corresponds to the purely non-local case and $r=1$ to the purely local case.
}
\label{fig:SeebeckNLyL}
\end{figure}

The behavior of the local and non-local components of the Seebeck coefficient are shown in  Fig. \ref{fig:SeebeckNLyL}  as functions of $\theta$ for different values of the factor $r$. 
We see the high sensitivity of the non-local thermoelectric effect to the angle $\theta$, in contrast with the local one, which depends mildly on this angle. 
This figure highlights the importance of implementing a symmetric temperature bias ($r=-1$) to cleanly observe the non-local thermoelectric effect.

\section{\label{sec:Conclusions}Summary and Conclusions}
We have shown the existence of a nonlocal thermoelectric response in a superconducting wire hosting SOC with twisted orientations of a magnetic field with respect to the wire SOC main axis. 
We focused on the linear response regime, corresponding to a small temperature bias.
We predict this effect to take place in systems akin to those typically used in the search for Majorana zero modes \cite{mourik2012signatures,deng2016majorana,chen2017experimental,PhysRevLett.123.107703,nichele2017scaling}. 

The possible impact of Majorana zero modes in  thermoelectric effects has been explored in structures with quantum dots \cite{Lopez2014May,Leijnse2014Jan,Valentini2015,Chi2020Dec,Hou2013Aug,
Ramos-Andrade2016Oct,Majek2022Feb,Buccheri2022Feb,Smirnov2023Apr,Wang2019May,Klees2023Jun}.
 In contrast, the non-local effect addressed here is related to the emergence of Bogoliubov Fermi points. This takes place when the gap of the topological phase is partially closed by a twist of the magnetic field with respect to the SOC, beyond a critical alignment and has been recently observed in two-dimensional samples of these materials \cite{phan2022detecting}.
 
 The estimate of the Seebeck coefficient, albeit small, is compatible with measured thermovoltages in other systems \cite{Molenkamp1990Aug,Dutta2019Jan,PhysRevApplied.14.034019}, assuming temperature differences of $10-100$~mK. Its behavior is strongly sensitive to the relative orientation of the magnetic field and the SOC, providing a valuable hallmark of this fundamental property.

\section{\label{sec:Acknow}Acknowledgements}
L.T. acknowledges the Georg Forster Fellowship from the Humboldt Stiftung. L.A., L.T. and J.H.M. are gratefule for 
support from CONICET as well as FonCyT, Argentina, through Grants No.  PICT-2018-04536 and No. PICT 2020-A-03661.  L.A. would also like to thank the Institut Henri Poincar\'e (UAR 839 CNRS-Sorbonne Universit\'e) and the LabEx CARMIN (ANR-10-LABX-59-01) for their support. A.B. and F.T. acknowledge the MUR-PRIN 2022 - Grant No. 2022B9P8LN - (PE3)-Project NEThEQS  ''Non-equilibrium coherent thermal effects in quantum systems'' in PNRR Mission 4 - Component 2 - Investment 1.1 ''Fondo per il Programma Nazionale di Ricerca e Progetti di Rilevante Interesse Nazionale (PRIN)'' funded  by the European Union - Next Generation EU and the Royal Society through the International Exchanges between the UK and Italy (Grants No. IEC R2 192166). 
A.B. acknowledges EU's Horizon 2020 Research and Innovation Framework Programme Grant No. 964398 (SUPERGATE) and No. 101057977 (SPECTRUM),  and CNR project QTHERMONANO.

%%%%%%%%%%%%%%%%%%%%%%%%%%%%%%%%%%%%%%%%%%%%%%%%%%
%%%%%%%%%%%%%%%%% %Appendix %%%%%%%%%%%%%%%%%%%%%%
%%%%%%%%%%%%%%%%%%%%%%%%%%%%%%%%%%%%%%%%%%%%%%%%%%
\appendix

\section{\label{sec:GreenCalculations}Calculation of Green's functions}\label{green}
We present here the Dyson's equations leading to the calculation of the retarded Green's functions. 
\subsection{Retarded/advanced}
The Dyson equation for the Fourier-transformed
retarded Green's functions, ${\bf G}^r_{\alpha,\gamma}(t,t')=-i \theta (t-t') \langle \left\{\psi_\alpha(t),\psi^{\dagger}_\gamma(t')\right\}\rangle$ corresponding to the different Nambu spinors
$\psi_{\alpha,\gamma}(t)\equiv {\bf c}_j(t), {\bf b}(t), {\bf d}(t)$ reads
\begin{eqnarray}\label{dyson}
    {\bf G}^r_{d0}(\varepsilon)&=&  {\bf G}^r_{dd}(\varepsilon){\bf t}_{\rm d} {\bf g}^r_{00}(\varepsilon),  \\
        {\bf G}^r_{dd}(\varepsilon)&=& \tilde{\bf g}^r_{dd}(\varepsilon) + {\bf G}^r_{d0}(\varepsilon) {\bf t}_{\rm d} \tilde{\bf g}^r_{dd}(\varepsilon)
        +{\bf G}^r_{dN}(\varepsilon) {\bf t}_{\rm N} \tilde{\bf g}^r_{dd}(\varepsilon), \nonumber \\
               {\bf G}^r_{dN}(\varepsilon)&=&  {\bf G}^r_{dd}(\varepsilon){\bf t}_{\rm N} {\bf g}^r_{NN}(\varepsilon), \nonumber \\
                {\bf G}^r_{Nd}(\varepsilon)&=& {\bf g}^r_{NN}(\varepsilon) {\bf t}_{\rm N} {\bf G}^r_{dd}(\varepsilon),
        \nonumber
\end{eqnarray}
where we have introduced the definition of the retarded Green's function of the quantum dot isolated from the rest of the subsystems, 
\begin{equation}
\tilde{\bf g}^r_{dd}(\varepsilon)=\left[\varepsilon \tau^0\otimes \sigma^0 + \varepsilon_{\rm d} \tau^z \otimes \sigma^0 +
B_{\rm d} \tau^0 \otimes \vec{n}_{B} \cdot \vec{\sigma} \right]^{-1},
\end{equation}
as well as 
the Green's function of the wire connected to the two superconducting reservoirs but disconnected from the quantum dot and the normal lead,
evaluated at the connecting site $0$. It reads 
\begin{equation}
{\bf g}^r_{00}(\varepsilon)=\left[\tilde{\bf g}^r(\varepsilon)^{-1}-\boldsymbol{\Sigma}^r_{1}(\varepsilon)-
\boldsymbol{\Sigma}^r_{2}(\varepsilon)\right]^{-1}|_{00},
\end{equation}
where $\tilde{\bf g}^r(\varepsilon)$ the Green's function of the free wire.
Substituting in Eqs. (\ref{dyson}) the first equation in the second one we get
\begin{equation}
    {\bf G}^r_{dd}(\varepsilon) = \left[ \tilde{\bf g}^r_{dd}(\varepsilon)^{-1} - \boldsymbol{\Sigma}^r_{\rm S}(\varepsilon)-
    \boldsymbol{\Sigma}^r_{\rm N}(\varepsilon)\right]^{-1}.
\end{equation}
We have also introduced 
the self-energies
\begin{eqnarray}
    \boldsymbol{\Sigma}^r_{\rm S}(\varepsilon) & = & {\bf t}_{\rm d} {\bf g}^r_{00}(\varepsilon){\bf t}_{\rm d} , \nonumber \\
    \boldsymbol{\Sigma}^r_{\rm N}(\varepsilon) & = & {\bf t}_{\rm N} \tilde{\bf g}^r_{NN}(\varepsilon){\bf t}_{\rm N},
\end{eqnarray}
 (notice that this is a $4(N_{\rm w}+1)\times 4(N_{\rm w}+1)$ matrix), while $\boldsymbol{\Sigma}^r_{j}(\varepsilon), \; j=L,R$
are the self-energies describing the coupling of the wire to the superconducting leads $L,R$. These can be also represented as 
$4(N_{\rm w}+1) \times 4(N_{\rm w}+1)$ matrices with non-vanishing $4\times 4$ submatrices associated with the spacial coordinates
$-N_{\rm w}/2$ (for $j=L$) and $N_{\rm w}/2$ (for $j=R$), respectively. 
The non-vanishing  self-energy matrices read
\begin{equation}
 \boldsymbol{\Sigma}^r_{j}(\varepsilon) = {\bf t}_j \tilde{\bf g}^r_{S}(\varepsilon){\bf t}^\dagger_j,
%  \boldsymbol{\Sigma}^r_{j}(\varepsilon) = {\bf t}_j(\phi) \tilde{\bf g}^r_{S}(\varepsilon){\bf t}^\dagger_j(\phi), 
\end{equation}
where ${\bf t}_j$ is the matrix element representing the contact between the wire and the reservoirs. This contains the hopping as well as
the spin-orbit terms
and $\tilde{\bf g}^r_{S}(\varepsilon)$ is the Green's function for the semi-infinite superconducting wire representing the reservoir. This is calculated by means of a recursive algorithm \cite{lopezsancho}.

The self-energy describing the contact to the normal lead reads $\boldsymbol{\Sigma}^r_{\rm N}(\varepsilon) = t_{\rm N}^2 \tilde{\bf g}^r_{N}(\varepsilon)$,
where $\tilde{\bf g}^r_{N}$ the Green's function  of the normal lead, which is also calculated by a recursive algorithm.

The advanced functions can be calculated from
\begin{equation}
{\bf G}_{ij}^a(\varepsilon)= \left[{\bf G}^r_{ji}(\varepsilon)\right]^{\dagger}.
\end{equation}

\subsection{Lesser}
The lesser Green's functions can be calculated from the retarded/advanced ones by recourse to Langreth's rule:
\begin{eqnarray}\label{lesser1}
    {\bf G}^<_{d0}(\varepsilon)&=&  {\bf G}^<_{dd}(\varepsilon){\bf t}_{\rm d} {\bf g}^a_{00}(\varepsilon)+
    {\bf G}^r_{dd}(\varepsilon){\bf t}_{\rm d} {\bf g}^<_{00}(\varepsilon),  \nonumber \\
        {\bf G}^<_{dN}(\varepsilon)&=&  {\bf G}^<_{dd}(\varepsilon){\bf t}_{\rm N} {\bf g}^a_{NN}(\varepsilon)+
    {\bf G}^a_{dd}(\varepsilon){\bf t}_{\rm N} {\bf g}^<_{NN}(\varepsilon),  \nonumber \\
        {\bf G}^<_{dd}(\varepsilon)&=& {\bf G}^r_{dd}(\varepsilon)\left[ \boldsymbol{\Sigma}^<_{\rm S}(\varepsilon)+ \boldsymbol{\Sigma}^<_{\rm N}
        \right] {\bf G}^a_{dd}(\varepsilon),\nonumber\\
        {\bf g}^<_{00}(\varepsilon) & = &    {\bf g}^r_{0, -N_{\rm w}/2}(\varepsilon) \boldsymbol{\Sigma}^<_{L}(\varepsilon) {\bf g}^a_{-N_{\rm w}/2,0}(\varepsilon) \nonumber \\
       & +& {\bf g}^r_{0, N_{\rm w}/2}(\varepsilon) \boldsymbol{\Sigma}^<_{R}(\varepsilon){\bf g}^a_{N_{\rm w}/2,0}(\varepsilon).
\end{eqnarray}
The different self-energies are
\begin{eqnarray}\label{lesser-def}
    \boldsymbol{\Sigma}^<_{\alpha}(\varepsilon) &=& {\bf f}_{\alpha}(\varepsilon)\left[ \boldsymbol{\Sigma}^a_{\alpha}(\varepsilon)-
    \boldsymbol{\Sigma}^r_{\alpha}(\varepsilon)\right] , \; \alpha=L,R,{\rm N},\nonumber \\
    \boldsymbol{\Sigma}^<_{\rm S}(\varepsilon) &=& \sum_{j=L,R} \boldsymbol{\Lambda}_j^r(\varepsilon)\boldsymbol{\Sigma}^<_{j}(\varepsilon) 
    \boldsymbol{\Lambda}_j^a(\varepsilon),
\end{eqnarray}
where
\begin{eqnarray}
    \boldsymbol{\Lambda}_1^{r}(\varepsilon)&=&{\bf t}_d  {\bf g}^r_{0, -N_{\rm w}/2}(\varepsilon),\nonumber \\
    \boldsymbol{\Lambda}_2^{r}(\varepsilon)&=&{\bf t}_d  {\bf g}^r_{0, N_{\rm w}/2}(\varepsilon),
\end{eqnarray}
with $\boldsymbol{\Lambda}_j^{a}(\varepsilon)=\left[\boldsymbol{\Lambda}_j^{r}(\varepsilon)\right]^{\dagger}$.

Substituting we get
\begin{eqnarray}\label{lesser2}
        {\bf G}^<_{dd}(\varepsilon)&=& \sum_{j=L,R}{\bf \cal G}^r_{dj}(\varepsilon) \boldsymbol{\Sigma}^<_{j}(\varepsilon){\bf \cal G}^a_{jd}(\varepsilon)
        \nonumber \\
        &+& {\bf G}^r_{dd}(\varepsilon)
 \boldsymbol{\Sigma}^<_{\rm N} {\bf G}^a_{dd}(\varepsilon),
\end{eqnarray}
where
\begin{equation}
    {\bf \cal G}^r_{dj}(\varepsilon) = {\bf G}^r_{dd}(\varepsilon)\boldsymbol{\Lambda}_j^{r}(\varepsilon)=\left[ {\bf \cal G}^a_{jd}(\varepsilon)\right]^\dagger.
\end{equation}

\begin{figure}[h]
	\centering
	\includegraphics[width=\columnwidth]{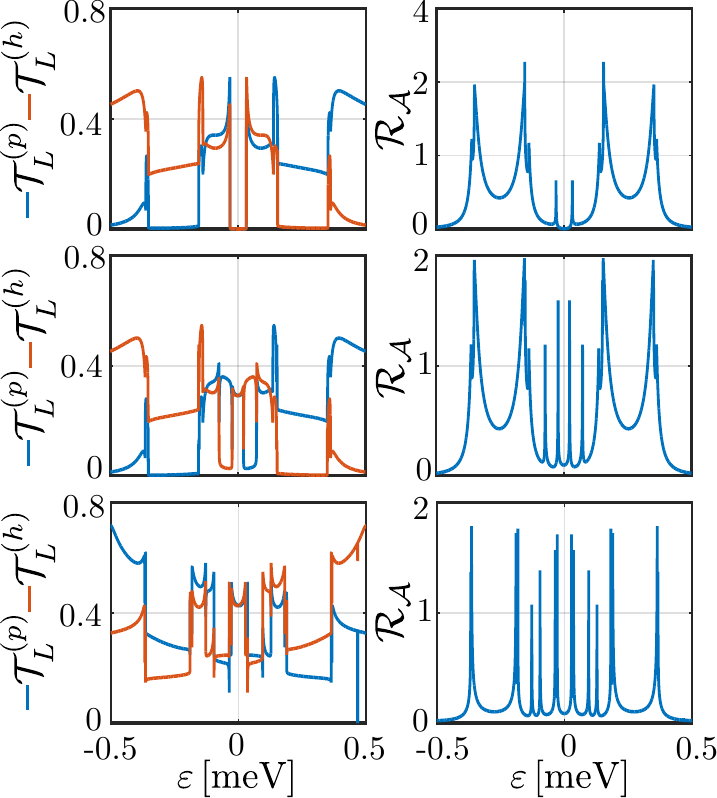}
	\caption{Transmissions and Andreev reflection functions for a system with $t=5\,\rm{meV}$, $\Delta_B=0.5\,\rm{meV}$, $\lambda=0.25\,\rm{meV}$, $\Delta=0.2\,\rm{meV}$, $t_d=2.5\,\rm{meV}$ and: (a) $\mu=-9.9\,{\rm{meV}}$, $\theta =\pi/2$; (b) $\mu=-9.9\,{\rm{meV}}$, $\theta =\pi/4$ or (c) $\mu=-8.5\,{\rm{meV}}$, $\theta =\pi/4$.
}
	\label{fig:TL-Andreev}
\end{figure}

\begin{figure}[h]
	\centering
	\includegraphics[width=\columnwidth]{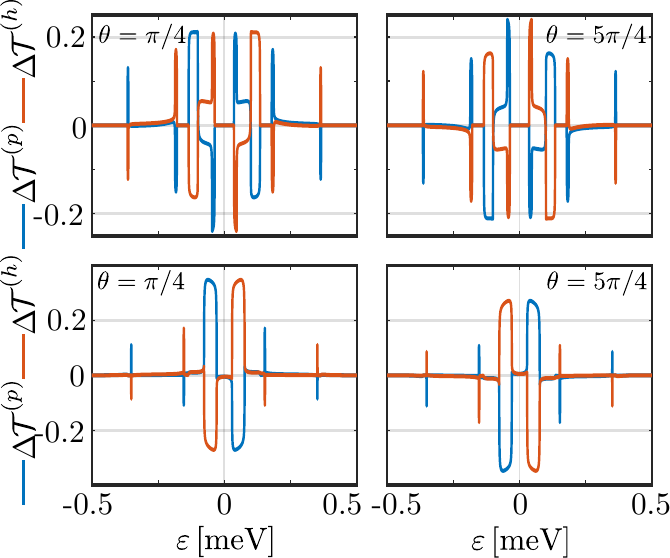}
	\caption{Difference in transmission functions, $\Delta{\cal{T}}^{(p)}={\cal{T}}_L^{(p)}-{\cal{T}}_R^{(p)}$ and $\Delta{\cal{T}}^{(h)}={\cal{T}}_L^{(h)}-{\cal{T}}_R^{(h)}$ involved in the calculation of ${\cal{L}}^{\rm nl}_{eq}$ for a system with $t=5\,\rm{meV}$, $\Delta_B=0.5\,\rm{meV}$, $\lambda=0.25\,\rm{meV}$, $\Delta=0.2\,\rm{meV}$, $t_d=2.5\,\rm{meV}$ and top: $\mu=-9.9\,\rm m eV$; bottom: $\mu=-8.5\,\rm m eV$. \label{dife}
}
\end{figure}

\subsection{Identities}
The following identity can be shown
\begin{eqnarray}
    & & {\bf G}^a_{dd}(\varepsilon)-{\bf G}^r_{dd}(\varepsilon)={\bf G}^r_{dd}(\varepsilon)
    \left[\boldsymbol{\Sigma}^a_{T}(\varepsilon) -\boldsymbol{\Sigma}^r_{T}(\varepsilon)\right]{\bf G}^a_{dd}(\varepsilon), \nonumber \\
  & & \boldsymbol{\Sigma}^{r/a}_{T}(\varepsilon) =    \boldsymbol{\Sigma}^{r/a}_{\rm S}(\varepsilon) +
    \boldsymbol{\Sigma}^{r/a}_{\rm N}(\varepsilon).
\end{eqnarray}

Another important identity can be derived by noticing that the current defined in Eq. (\ref{current}) should be zero in equilibrium.
This implies
\begin{equation}
     2\mbox{Re} \left[ \boldsymbol{\Sigma}^r_N {\bf G}^<_{d,d}(\varepsilon) +\boldsymbol{\Sigma}^<_N {\bf G}^a_{d,d}(\varepsilon) \right]=0.
\end{equation}
Substituting the definitions of  all these quantities we get

\begin{multline}\label{iden}
  f_N  \boldsymbol{\Gamma}_{N}(\varepsilon)  \left[ {\bf G}^a_{d,d}(\varepsilon) 
  - {\bf G}^r_{d,d}(\varepsilon)\right]-f_N \boldsymbol{\Gamma}_{N}(\varepsilon)\times \\
            \left\{
            \sum_{j=1,2}{\bf \cal G}^r_{dj}(\varepsilon) \boldsymbol{\Gamma}_{j}(\varepsilon){\bf \cal G}^a_{jd}(\varepsilon)+
            {\bf G}^r_{dd}(\varepsilon)    \boldsymbol{\Gamma}_{N}(\varepsilon)  
 {\bf G}^a_{dd}(\varepsilon)\right\} =0,
\end{multline}
%\begin{eqnarray}\label{iden}
%& & f_N  \left[ \boldsymbol{\Sigma}^a_N -\boldsymbol{\Sigma}^r_N \right]  \left[ {\bf G}^a_{d,d}(\varepsilon) 
%  - {\bf G}^r_{d,d}(\varepsilon)\right]\nonumber \\
%  & &+f_N \left[ \boldsymbol{\Sigma}^r_N - \boldsymbol{\Sigma}^a_N\right]\times \nonumber \\
%          & & \left\{
%            \sum_{j=1,2}{\bf \cal G}^r_{dj}(\varepsilon) \left[\boldsymbol{\Sigma}^a_{j}(\varepsilon)-\boldsymbol{\Sigma}^r_{j}(\varepsilon)
%            \right]{\bf \cal G}^a_{jd}(\varepsilon)+
%            \right. \nonumber \\
%          & &\left.+{\bf G}^r_{dd}(\varepsilon)    \left[ \boldsymbol{\Sigma}^a_N -\boldsymbol{\Sigma}^r_N \right]  
% {\bf G}^a_{dd}(\varepsilon)\right\} =0,
%  \end{eqnarray}
where $f_N $ is the Fermi-Dirac distribution function corresponding to the equilibrium system. Since this function is a common factor in all the terms,
  this identity is zero for any argument of $f_N$. We have introduced the definition
  $\boldsymbol{\Gamma}_{j}(\varepsilon)=\boldsymbol{\Sigma}^a_{j}(\varepsilon)-\boldsymbol{\Sigma}^r_{j}(\varepsilon)$, with $j=L,R,N$.

\section{\label{sec:DetailsOfJ}Details on the calculation of $J$}
Using Eqs.~(\ref{lesser1}) we can rewrite the argument of Eq.~(\ref{current}) as follows
%\begin{widetext}
\begin{eqnarray}
   {\bf t}_{\rm N}{\bf G}^<_{Nd}(\varepsilon) &=&  {\bf t}_{\rm N}{\bf g}^<_{NN}(\varepsilon) {\bf t}_{\rm N} {\bf G}^a_{dd}(\varepsilon) 
   \nonumber \\
   &+& 
   {\bf t}_{\rm N}{\bf g}^r_{NN}(\varepsilon) {\bf t}_{\rm N} {\bf G}^<_{dd}(\varepsilon)
   \nonumber \\
   &=&  
   \boldsymbol{\Sigma}^<_N {\bf G}^a_{dd}(\varepsilon) 
   + \boldsymbol{\Sigma}^r_N{\bf G}^<_{dd}(\varepsilon).
\end{eqnarray}
%\end{widetext}

Similarly, using Eqs. (\ref{lesser-def}) and (\ref{lesser2}) we can write
%\begin{widetext}
%\begin{multline}
%     2 \mbox{Re}\left[ \boldsymbol{\Sigma}^<_N %{\bf G}^a_{dd}(\varepsilon) \right] =
%  {\bf f}_N   \boldsymbol{\Gamma}_{N}(\varepsilon)  \left[ {\bf G}^a_{dd}(\varepsilon) 
%  - {\bf G}^r_{dd}(\varepsilon)\right] 
%  \\ 
%    =\left[{\bf f}_N  -f^+_N {\bf 1} \right]\boldsymbol{\Gamma}_{N}(\varepsilon)  \left[ {\bf G}^a_{dd}(\varepsilon) 
%  - {\bf G}^r_{dd}(\varepsilon)\right] 
%   %\\
%   +f^+_N  \boldsymbol{\Gamma}_{N}(\varepsilon)  \left[ {\bf G}^a_{dd}(\varepsilon) 
 % - {\bf G}^r_{dd}(\varepsilon)\right],
%\end{multline}
%\end{widetext}

%\begin{widetext}
%\begin{align}
%     2 \mbox{Re}\left[ \boldsymbol{\Sigma}^<_N {\bf G}^a_{dd}(\varepsilon) \right] &=
%  {\bf f}_N   \boldsymbol{\Gamma}_{N}(\varepsilon)  \left[ {\bf G}^a_{dd}(\varepsilon) 
%  - {\bf G}^r_{dd}(\varepsilon)\right] 
%  \\ \nonumber
%    &=\left[{\bf f}_N  -f^+_N {\bf 1} \right]\boldsymbol{\Gamma}_{N}(\varepsilon)  \left[ {\bf G}^a_{dd}(\varepsilon) 
%  - {\bf G}^r_{dd}(\varepsilon)\right] 
%   %\\
%   +f^+_N  \boldsymbol{\Gamma}_{N}(\varepsilon)  \left[ {\bf G}^a_{dd}(\varepsilon) 
%  - {\bf G}^r_{dd}(\varepsilon)\right],
%\end{align}
%\end{widetext}
%

\begin{widetext}
\vspace{+0mm}
\begin{align}
     2 \mbox{Re}\left[ \boldsymbol{\Sigma}^<_N {\bf G}^a_{dd}(\varepsilon) \right] &=
  {\bf f}_N   \boldsymbol{\Gamma}_{N}(\varepsilon)  \left[ {\bf G}^a_{dd}(\varepsilon) 
  - {\bf G}^r_{dd}(\varepsilon)\right] 
  \\ \nonumber
    &=\left[{\bf f}_N  -f^+_N {\bf 1} \right]\boldsymbol{\Gamma}_{N}(\varepsilon)  \left[ {\bf G}^a_{dd}(\varepsilon) 
  - {\bf G}^r_{dd}(\varepsilon)\right] 
   %\\
   +f^+_N  \boldsymbol{\Gamma}_{N}(\varepsilon)  \left[ {\bf G}^a_{dd}(\varepsilon) 
  - {\bf G}^r_{dd}(\varepsilon)\right],
\end{align}
\vspace{+0mm}
\begin{align}
2\mbox{Re}\left[ \boldsymbol{\Sigma}^r_N {\bf G}^<_{dd}(\varepsilon) \right] %=\nonumber\\
           =\,\,&\boldsymbol{\Gamma}_{N}(\varepsilon)  \left\{
            \sum_{j=L,R}      \left( f_j -f^+_N\right){\bf \cal G}^r_{dj}(\varepsilon) \boldsymbol{\Gamma}_{j}(\varepsilon){\bf \cal G}^a_{jd}(\varepsilon)
            +f^+_N\sum_{j=L,R}{\bf \cal G}^r_{dj}(\varepsilon) \boldsymbol{\Gamma}_{j}(\varepsilon){\bf \cal G}^a_{jd}(\varepsilon)+\right.\\ \nonumber
%            \nonumber \\
            &\left.+ {\bf G}^r_{dd}(\varepsilon) \left[{\bf f}_N  -f^+_N {\bf 1} \right]  \boldsymbol{\Gamma}_{N}(\varepsilon)  
 {\bf G}^a_{dd}(\varepsilon)
 %\nonumber\\ 
 +{\bf G}^r_{dd}(\varepsilon) f^+_N   \boldsymbol{\Gamma}_{N}(\varepsilon)  
 {\bf G}^a_{dd}(\varepsilon)\right\}.
\end{align}
\vspace{+5mm}
\end{widetext}

%\begin{eqnarray}
%     2 \mbox{Re}& &\left[ \boldsymbol{\Sigma}^<_N {\bf G}^a_{dd}(\varepsilon) \right] =
 % {\bf f}_N   \left[ \boldsymbol{\Sigma}^a_N -\boldsymbol{\Sigma}^r_N \right]  \left[ {\bf G}^a_{dd}(\varepsilon) 
%  - {\bf G}^r_{dd}(\varepsilon)\right] \nonumber \\
%  & & 
%  \;\;\;\;\;\;\;\;
%  =\left[{\bf f}_N  -f^+_N {\bf 1} \right]\left[ %\boldsymbol{\Sigma}^a_N -\boldsymbol{\Sigma}^r_N %\right]  \left[ {\bf G}^a_{dd}(\varepsilon) 
%  - {\bf G}^r_{dd}(\varepsilon)\right] \nonumber \\
%   & &
%   \;\;\;\;\;\;\;\;\;\;\;\; 
%   +f^+_N  \left[ \boldsymbol{\Sigma}^a_N -\boldsymbol{\Sigma}^r_N \right]  \left[ {\bf %G}^a_{dd}(\varepsilon) 
%  - {\bf G}^r_{dd}(\varepsilon)\right],
%\end{eqnarray}
where
\begin{eqnarray}
    {\bf f}_{\alpha}(\varepsilon)&=& f_\alpha(\varepsilon) \tau^0\otimes \sigma^0,\;\;\alpha=L,R,\nonumber\\
    {\bf f}_{N}(\varepsilon)&=& \left(\begin{array}{cc}
    f_N^+(\varepsilon) \sigma^0 &0\\
    0 & f_N^-(\varepsilon) \sigma^0
    \end{array}\right).
\end{eqnarray}

Calculating the sum over the elements of these matrices, we notice that $\sum_{\ell=1,2} \left[{\bf f}_N  -f^+_N {\bf 1} \right] =0$ and $\sum_{\ell=3,4} \left[{\bf f}_N  -f^-_N {\bf 1} \right] =0$.

On the other hand, using the identity of Eq. (\ref{iden}) we get
\begin{widetext}
\vspace{+0mm}
\begin{eqnarray}
    & &  2\mbox{Re} \left[ \boldsymbol{\Sigma}^r_N {\bf G}^<_{dd}(\varepsilon) +\boldsymbol{\Sigma}^<_N {\bf G}^a_{dd}(\varepsilon) \right]
    = \left[{\bf f}_N  -f^+_N {\bf 1} \right]\boldsymbol{\Gamma}_{N}(\varepsilon)\times 
     %\\  
     \left[ {\bf G}^a_{dd}(\varepsilon) 
  - {\bf G}^r_{dd}(\varepsilon)\right] \nonumber\\
  & &  -\boldsymbol{\Gamma}_{N}(\varepsilon) \times \left\{
           \sum_{j=L,R}  \left( f_j -f^+_N\right) {\bf \cal G}^r_{dj}(\varepsilon) \boldsymbol{\Gamma}_{j}(\varepsilon){\bf \cal G}^a_{jd}(\varepsilon)
           +{\bf G}^r_{dd}(\varepsilon) \left[{\bf f}_N  -f^+_N {\bf 1} \right]  \boldsymbol{\Gamma}_{N}(\varepsilon)  
 {\bf G}^a_{dd}(\varepsilon) \right\}.
\end{eqnarray}
\vspace{-0mm}
\end{widetext}

%\begin{eqnarray}
%     & & 2\mbox{Re} \left[ \boldsymbol{\Sigma}^r_N {\bf G}^<_{dd}(\varepsilon) +\boldsymbol{\Sigma}^<_N {\bf G}^a_{dd}(\varepsilon) \right]\nonumber \\
%     & &= \left[{\bf f}_N  -f^+_N {\bf 1} \right]\left[ \boldsymbol{\Sigma}^a_N -\boldsymbol{\Sigma}^r_N \right]  \left[ {\bf G}^a_{dd}(\varepsilon) 
%  - {\bf G}^r_{dd}(\varepsilon)\right] \nonumber \\
%  & & +\left[ \boldsymbol{\Sigma}^r_N - \boldsymbol{\Sigma}^a_N\right]\times \nonumber \\
%          & & + \left\{
%           \sum_{j=L,R}  \left( f_j -f^+_N\right) {\bf \cal G}^r_{dj}(\varepsilon) \left[\boldsymbol{\Sigma}^a_{j}(\varepsilon)-\boldsymbol{\Sigma}^r_{j}(\varepsilon)
%            \right]{\bf \cal G}^a_{jd}(\varepsilon)+
%            \right. \nonumber \\
%         & &  \left.+ {\bf G}^r_{dd}(\varepsilon) \left[{\bf f}_N  -f^+_N {\bf 1} \right]  \left[ \boldsymbol{\Sigma}^a_N -\boldsymbol{\Sigma}^r_N \right]  
% {\bf G}^a_{dd}(\varepsilon) \right\} .
%\end{eqnarray}

%\textcolor{red}{The last two elements are treated in a similar way.}

\section{\label{sec:PropertiesTransmissionFunctions}Properties of the transmission functions}
The behavior of the particle and hole transmission functions defined in Eqs.~(\ref{tpj}) are illustrated in Fig. \ref{fig:TL-Andreev} for the left reservoir, along with the Andreev reflection for the same parameters. The functions corresponding to the other reservoir exhibit similar features.

We can verify that  these functions satisfy:
\begin{eqnarray}\label{1}
{\cal T}_j^{\rm (p)}(\varepsilon,\theta)&=&{\cal T}_j^{\rm (h)}(-\varepsilon,\theta),\nonumber \\
{\cal T}_L^{\rm (p)}(\varepsilon,\pm|\theta|)&=&{\cal T}_R^{\rm (p)}(\varepsilon,\pm(|\theta|+\pi/2)),\nonumber \\
{\cal T}_L^{\rm (h)}(\varepsilon,\pm|\theta|)&=&{\cal T}_R^{\rm (h)}(\varepsilon,\pm(|\theta|+\pi/2)),\nonumber \\
{\cal T}_L^{\rm (p)}(\varepsilon,\theta)&=&{\cal T}_R^{\rm (p)}(\varepsilon,\theta\pm\pi),\nonumber \\
{\cal T}_L^{\rm (h)}(\varepsilon,\theta)&=&{\cal T}_R^{\rm (h)}(\varepsilon,\theta\pm\pi).
\end{eqnarray}

In Fig. \ref{dife} we show the difference between the transmission functions associated with the $L$ and $R$ superconductors.
This combination of transmission functions determine the non-local thermoelectric response and illustrate the 
symmetry properties mentioned above.

%%%%%%%%%%%%%%%%%%%%%%%%%%%%%%%%%%%%%%%%%%%%%%%%%%
\bibliography{super-wire.bib}

\newpage

\end{document}